\documentclass[prd,aps,showpacs,nofootinbib]{revtex4}
\usepackage{amsmath}
\usepackage{amssymb}
\usepackage{amsfonts}
\usepackage{graphicx,bm}
\usepackage{dcolumn}
\usepackage{colordvi}
\usepackage{color}

\usepackage{color,amsxtra}
\usepackage{epsf}
\usepackage{enumerate}
\usepackage{hhline}
\usepackage{array}
\usepackage{tabularx}
\newcommand{\be}{\begin{equation}}
\newcommand{\ee}{\end{equation}}
\newcommand{\bea}{\begin{eqnarray}}
\newcommand{\eea}{\end{eqnarray}}
\newcommand{\beaa}{\begin{eqnarray*}}
\newcommand{\eeaa}{\end{eqnarray*}}

\newcommand{\abs}[1]{\vert{#1}\vert}

\def\be{\begin{equation}}
\def\ee{\end{equation}}
\def\bea{\begin{eqnarray}}
\def\eea{\end{eqnarray}}

\begin{document}


\title{No further gravitational wave modes in $F(T)$ gravity}
\author{Kazuharu Bamba$^{1,}$\footnote{
E-mail address: bamba@kmi.nagoya-u.ac.jp}, 
Salvatore Capozziello$^{1, 2, 3,}$\footnote{E-mail address: capozziello@na.infn.it}, \\
Mariafelicia De Laurentis$^{1, 2, 3,}$\footnote{E-mail address: felicia@na.infn.it}, 
Shin'ichi Nojiri$^{1, 4,}$\footnote{E-mail address:
nojiri@phys.nagoya-u.ac.jp} 
and 
Diego S\'{a}ez-G\'{o}mez$^{1, 5, 6,}$\footnote{
E-mail address: diego.saezgomez@uct.ac.za}
}
\affiliation{
$^1$Kobayashi-Maskawa Institute for the Origin of Particles and the
Universe,
Nagoya University, Nagoya 464-8602, Japan\\
$^2$Dipartimento di Fisica,
Universit\`{a} di Napoli {}``Federico II'' and\\ 
$^3$INFN Sez. di Napoli, Compl. Univ. di Monte S. Angelo, Edificio G, Via Cinthia, I-80126, Napoli, Italy\\
$^4$Department of Physics, Nagoya University, Nagoya 464-8602, Japan\\ 
$^5$Astrophysics, Cosmology and Gravity Centre (ACGC) and \\ 
Department of Mathematics and Applied Mathematics, University of Cape Town, Rondebosch 7701, Cape Town, South Africa \\
$^6$Fisika Teorikoaren eta Zientziaren Historia Saila, Zientzia eta Teknologia Fakultatea,\\
Euskal Herriko Unibertsitatea, 644 Posta Kutxatila, 48080 Bilbao, Spain.
}


\begin{abstract}

We explore the possibility of further gravitational wave modes  in $F(T)$ gravity,   where $T$ is the torsion scalar in teleparallelism.
It is explicitly demonstrated that gravitational wave modes in $F(T)$ gravity are equivalent to those in General Relativity. This result is achieved by calculating the Minkowskian limit for a class of analytic  function of $F(T)$. 
This consequence is also confirmed by the preservative analysis 
around the flat background in the weak field limit 
with the scalar-tensor representation of $F(T)$ gravity. 

\end{abstract}


\maketitle

\textit{Introduction}-- 
It has been suggested from recent precise cosmological observations~\cite{SN1, LSS, Eisenstein:2005su, WMAP, Jain:2003tba} 
that if the current universe is considered to strictly be homogeneous, 
there exist not only dark matter and baryons 
but also the so-called "dark energy", whose contribute provides a negative pressure able to accelerate the universe today. It is not well understood yet whether dark energy can be described in terms of unknown contributions (such as cosmological constant, scalar field, some exotic fluid) 
or it has geometric origin, namely, 
gravity should be extended and modified at infrared limit on large distances. 
(for reviews about dark energy and extended/modified gravity theories, 
see, for example, Ref.~\cite{R-DE} and Ref.~\cite{F(R)-gravity}, 
respectively). 

In order to solve this dichotomy, one needs an {\it experimentum crucis} capable of discriminating between the two approaches. This signature could be related, on one hand, to the detection of new particles.
On the other hand, 
 supposing that gravitational waves are directly detected, 
by analyzing gravitational wave modes in Extended Theories of Gravity \cite{F(R)-gravity},  
which can explain the current cosmic acceleration, 
and those in General Relativity 
and comparing those modes with the observations, 
it is possible to judge, in principle, whether the origin of dark energy is 
geometric or not, that is, gravitation is described by 
Extended Theories of Gravity. 

Gravitational waves offer a remarkable opportunity to see the universe from a new perspective, providing an access to astrophysical insights that are available in no other way. Theoretical and experimental studies have been developed to understand the mechanisms for the production of gravitational waves, both in astrophysics and in cosmology~\cite{SS, Maggiore}. But even today, have not fully been resolved many conceptual problems and technical issues related to the production of gravitational waves from self-gravitating systems. In particular, Extended Theories of Gravity seem a viable scheme to overcome shortcomings related to infrared and ultraviolet behaviors of the gravitational field~\cite{F(R)-gravity}. Apart from the fundamental physics reasons, Extended Theories of Gravity have taken hold in cosmology, thanks to the fact that they ``naturally'' exhibit inflationary behaviors and therefore are able to overcome the shortcomings of Standard Cosmological Model~\cite{Kolb and Turner, W-C}. The cosmological models are capable of matching with the observations~\cite{F(R)-gravity}. 
There have been proposed the ways of constraining 
Extended Theories of Gravity as well as cosmological models 
in several model independent manners by using cosmological observational 
data, e.g., in Refs.~\cite{Aviles:2012ay, Visser:2003vq, Luongo:2011zz, Neben:2012wc}. 
These theories, from astrophysical and cosmological point of view, do not require the presence of candidates for dark energy and dark matter at fundamental level. The approach is very conservative taking into account only the gravity, radiation and baryonic matter, i.e., only the ``actually observed'' ingredients. In other words,  it is in full agreement with the early spirit of General Relativity which could not act in the same way at all scales.
In fact, General Relativity has successfully been probed in the weak-field limit (e.g., the Solar System experiments) and also in this case there is room for Extended Theories of Gravity which are not at all ruled out. In particular, it is possible to show that several models could satisfy both cosmological and the Solar System tests and could give rise to new effects~\cite{cap}.

As a candidate of Extended Theories of Gravity to realize 
the accelerated expansion of the universe, 
``teleparallelism''~\cite{Teleparallelism} has recently been considered, 
where the torsion scalar $T$ constructed by the Weitzenb\"{o}ck connection 
describes the action. 
In this theory, curvature becomes zero, although in General Relativity, 
the action is expressed by the Ricci scalar $R$ consisting of the Levi-Civita connection. 
It is known that if teleparallelism is extended, namely, 
the action is written by $F(T)$, which is a function of $T$, 
in the equivalent spirit to $F(R)$ gravity~\cite{Models-of-F(R),starobinsky} 
in General Relativity, 
inflation~\cite{Inflation-F-F} and/or the late time accelerated expansion 
of the universe~\cite{Bengochea:2008gz, Linder:2010py, WY-BGLL-BGL, 
Aviles:2013nga, A-C-F-G, BNO-BDJO-B} can be realized in such theories. 

An advantage of $F(T)$ gravity is that the order of the gravitational field equation is second, as that in General Relativity, while that is four in metric 
$F(R)$ gravity, so that the analysis of the cosmological expansion of the 
universe in $F(T)$ gravity can be much easier than that in $F(R)$ gravity. 
On the other hand, there exist some theoretical issues of 
$F(T)$ gravity, for example, 
conformal transformation~\cite{Yang:2010ji}, 
conformal invariance of teleparallelism~\cite{Maluf:2011kf}, 
conformal symmetry~\cite{Bamba:2013XX}, 
the local Lorentz invariance~\cite{LLI}, 
cosmological perturbation theories~\cite{CPT-F(T)}, 
a non-linear property of $F(T)$ gravity~\cite{Ong:2013qja}, 
superluminal modes~\cite{Izumi:2013dca}, 
and pure teleparallelism with non-minimal coupling to 
a scalar field~\cite{GLSW-GLS-XSL} that make such models very interesting to be studied.

In this Letter, we investigate gravitational waves in $F(T)$ gravity. 
It is explicitly shown that the gravitational wave modes in $F(T)$ gravity are equivalent to those in General Relativity. 
It should be emphasized that in Ref.~\cite{Li:2011rn}, 
the number of degrees of freedom in $F(T)$ gravity has been examined, 
and found that for four dimensional specetime, 
there exist three extra degrees of freedom, i.e., 
a massive vector field with a scalar field or 
a massless vector field with a scalar field. 
This implies that gravitational wave modes in $F(T)$ gravity 
are equivalent to those in General Relativity since such further modes do not contribute to the gravitational radiation in the post-Minkowskian limit. 
We also verify this consequence 
by using the preservative analysis in the weak filed limit 
around the flat background in 
the scalar-tensor representation of $F(T)$ gravity. 
Thus, our main result is consistent with that in Ref.~\cite{Li:2011rn} 
and the analysis of perturbations in $F(T)$ gravity executed 
in Ref.~\cite{Izumi:2012qj}. 
We use units  $k_\mathrm{B} = c = \hbar = 1$ and denote the
gravitational constant $8 \pi G$ by
${\kappa}^2 \equiv 8\pi/{M_{\mathrm{Pl}}}^2$
with the Planck mass of $M_{\mathrm{Pl}} = G^{-1/2} = 1.2 \times 
10^{19}$\,\,GeV. 


\textit{Teleparallelism}-- 
We introduce orthonormal tetrad components, i.e., the vierbein field, 
$h_A (x^{\mu})$ in teleparallelism, 
where at each point $x^{\mu}$ of a manifold, 
an index $A = 0, 1, 2, 3$ is used for the tangent space. 
Moreover, we have the relation between the metric and 
the vierbein field as 
$g_{\mu\nu}=\eta_{A B} h^A_\mu h^B_\nu$. 
Here, $\mu, \nu = 0, 1, 2, 3$ denote 
coordinate indices on the manifold, so that 
$h_A^\mu$ can be the tangent vector of the manifold. 
Also, the inverse of the vierbein field is derived by 
the relation $h^A_\mu h_A^\nu = \delta_\mu^\nu$. 
To construct a quantity to express the torsion of the space-time, 
we define the torsion tensor as 
$T^\rho_{\verb| |\mu\nu} \equiv 
\Gamma^\rho_{\verb| |\nu\mu} - \Gamma^\rho_{\verb| |\mu\nu} 
= h^\rho_A 
\left( \partial_\mu h^A_\nu - \partial_\nu h^A_\mu \right)$, 
where 
$\Gamma^\rho_{\verb| |\nu\mu} \equiv h^\rho_A \partial_\mu h^A_\nu$ 
is the Weitzenb\"{o}ck connection. 
By  the torsion tensor, we also construct the contorsion tensor 
$K^{\mu\nu}_{\verb|  |\rho} 
\equiv - \left(1/2\right) 
\left(T^{\mu\nu}_{\verb|  |\rho} - T^{\nu \mu}_{\verb|  |\rho} - 
T_\rho^{\verb| |\mu\nu}\right)$. 
Furthermore, using the torsion and contorsion tensors, we 
obtain the so-called superpotential 
$S_\rho^{\verb| |\mu\nu} \equiv \left(1/2\right) 
\left(K^{\mu\nu}_{\verb|  |\rho}+\delta^\mu_\rho \ 
T^{\alpha \nu}_{\verb|  |\alpha}-\delta^\nu_\rho \ 
T^{\alpha \mu}_{\verb|  |\alpha}\right)$. 
As a result, combining the torsion tensor and the superpotential leads to 
the torsion scalar 
$T \equiv S_\rho^{\verb| |\mu\nu} T^\rho_{\verb| |\mu\nu} = 
\left(1/4\right) T^{\rho\mu\nu}T_{\rho\mu\nu} + \left(1/2\right) 
T^{\rho\mu\nu}T_{\nu\mu\rho}-T_{\rho\mu}^{\verb|  |\rho}
T^{\nu\mu}_{\verb|  |\nu}$. 

In the above formulations of teleparallelism, 
we can write the Lagrangian density by the torsion scalar. 
We note that since the Weitzenb\"{o}ck connection is here adopted, 
the curvature does not appear and only torsion is present. 
This is the main difference from General Relativity, where 
we describe the Einstein-Hilbert action by the curvature scalar $R$. 
Along with the same spirit as in $F(R)$ gravity, 
we modify the action of teleparallelism as 
a function of $T$. 
Accordingly, we find~\cite{Linder:2010py} 
\begin{equation} 
S= \int d^4x \abs{h} \left( \frac{F(T)}{2{\kappa}^2} 
+{\mathcal{L}}_{\mathrm{M}} \right)\,. 
\label{eq:1}
\end{equation}
Here, we define 
$\abs{h} \equiv \det \left(h^A_\mu \right) = \sqrt{-g}$, 
where $g$ is the determinant of the metric $g_{\mu\nu}$, 
and 
${\mathcal{L}}_{\mathrm{M}}$ is the Lagrangian of matter. 
Variation of the action in Eq.~(\ref{eq:1}) leads to~\cite{Li:2011rn} 
$F^{\prime} \left[ 
\partial_\mu \left( h h_A^\nu S_\nu^{\verb| |\lambda\rho} \right) 
-h h_A^\rho S^{\mu\nu\lambda} T_{\mu\nu\rho} \right] 
+ F^{\prime \prime} h h_A^\nu S_\nu^{\verb| |\lambda\rho} \partial_\rho T 
+\left(1/2\right) h h_A^\lambda F(T) = {\kappa}^2 
{\mathcal{T}^{(\mathrm{M})}}_A^\lambda
$, where the prime denotes the derivative with respect to $T$ as 
$F^{\prime} \equiv \partial F/\partial T$ and 
$F^{\prime \prime} \equiv {\partial}^2 F/\partial T^2$ and 
${\mathcal{T}^{(\mathrm{M})}}_\rho^{\verb| |\nu}$ is 
the energy-momentum tensor of matter. 
The covariant description of the above gravitational field equation is 
\begin{equation} 
F^{\prime} \left( R_{\mu\nu} - \frac{1}{2} g_{\mu\nu} R \right) 
+ \frac{1}{2}g_{\mu\nu} \left[ F(T) - F^{\prime} T \right]
+ F^{\prime \prime} S_{\nu\mu\rho} \nabla^{\rho} T  = {\kappa}^2 
{\mathcal{T}^{(\mathrm{M})}}_{\mu\nu}\,, 
\label{eq:2}
\end{equation}
with $\nabla^{\rho}$ the covariant derivative 
and $R_{\mu\nu}$ and $R$ the Ricci tensor and Ricci scalar, 
respectively. We remark that for $F(T) = T$, this equation is 
equivalent to the Einstein equation. 
Moreover, the relation between $R$ and $T$ is 
$R = -T - 2{\nabla}^{\mu} \left( T^{\nu}_{\verb| |\mu\nu} 
\right)$ (see ~\cite{A-P}). 
This implies that  gravitational field equations
in General Relativity, described by the Einsten-Hilbert action, 
and its teleparallelism version, expressed by the torsion scalar $T$, 
are the same each other.

\textit{Gravitational waves in teleparallelism}-- 
Let us  investigate gravitational waves in teleparallelism 
by following discussions in Ref.~\cite{DeLaurentis:2011tp}. 
In order to obtain gravitational waves, the most natural starting point is to use linearized gravity. This means we adopt the weak field limit approximation~\cite{Maggiore, DeLaurentis:2011re}. The weak field limit is achieved by assuming that the space-time metric $g_{\mu\nu}$ is represented by the sum of the Minkowski space-time plus a small perturbation
\begin{eqnarray}
g_{\mu\nu}\,=\,\eta_{\mu\nu}+h_{\mu\nu}\,.
\label{metric}
\end{eqnarray}
Here, $h_{\mu\nu}$ is small and of first order ($ {\cal O}(h^2) \ll 1$). This implies that 
the gravitational field is required to be weak, and furthermore 
that the coordinate system is constrained to be approximately the Cartesian one. 
It is straightforward to  demonstrate that the relation (\ref{metric}) is equivalent to $g_{\mu\nu} = \eta_{ab}h^a_\mu h^b_\nu$
in the following way
\begin{eqnarray}\label{eq:metric}
g_{\mu\nu} &=& \eta_{ab}h^a_\mu h^b_\nu\,=\, \eta_{ab}\left(\delta^a_\mu+h^a_\mu\right)\left(\delta^b_\nu+h^b_\nu\right)\,=\,\eta_{ab}\left(\delta^a_\mu \delta^b_\nu+\delta^a_\mu h^b_\nu+h^a_\mu\delta^b_\nu+h^a_\mu h^b_\nu\right)\,=\,\nonumber\\&&
=  \eta_{ab}\delta^a_\mu \delta^b_\nu+\eta_{ab}\delta^a_\mu h^b_\nu+\eta_{ab}h^a_\mu\delta^b_\nu+\eta_{ab}h^a_\mu h^b_\nu\,=\,\nonumber\\&& 
=\eta_{\mu\nu}+\underbrace{\eta_{\mu b}h^b_\nu+ \eta_{a\nu}h^a_\mu}_{\simeq\, h_{\mu\nu}}+{\cal O}(h^2)\,,
\end{eqnarray}
where we only keep linear terms in $h_{\mu\nu}$ and higher order terms are discarded, since  we need to maintain the smallness of the perturbation.
In this sense, the perturbation on the vierbeins frame (tangent space) 
is connected to the metric perturbation in coordinates on the manifold. 
The metric perturbation, as well known, encapsulates gravitational waves, but 
contains additional non-radiative degrees of freedom as well. The metric perturbation $h_{\mu\nu}$ transforms as a tensor under the Lorentz transformations, but not under general coordinate transformations. We now compute all the quantities which are needed to describe linearized gravity. Hence, the Ricci tensor at first order of approximation in term of the perturbation is given by 
\begin{eqnarray}
R_{\mu\nu}^{(1)}\,
=\, \frac{1}{2}\left(\partial_\rho\partial_\nu {h^\rho}_\mu + \partial^\rho
\partial_\mu h_{\nu\rho} - \Box h_{\mu\nu} - \partial_\mu\partial_\nu h\right)\,,
\label{eq:ricci}
\end{eqnarray}
where $h = {h^\mu}_\mu$ is the trace of the metric perturbation, and $\displaystyle{\Box
= \partial_\rho\partial^\rho = \nabla^2 - \partial_t^2}$ is the wave 
operator. Here, $R_{\mu\nu}^{(n)}$ denotes the term in $R_{\mu\nu}$ that is of $n$-th order in $h_{\mu\nu}$. 
 Contracting once more, we find the scalar curvature 
\begin{eqnarray}
R^{(1)} = {R^\mu}_\mu = \partial_\rho\partial^\mu {h^\rho}_\mu - \Box h\,,
\label{eq:scalar}
\end{eqnarray}
and finally build the Einstein tensor 
\begin{eqnarray}
G_{\mu\nu} &=& R_{\mu\nu} - \frac{1}{2}\eta_{\mu\nu} R 
\nonumber\\
&=& \frac{1}{2}\left(\partial_\rho\partial_\nu {h^\rho}_\mu + \partial^\rho
\partial_\mu h_{\nu\rho} - \Box h_{\mu\nu} - \partial_\mu\partial_\nu h
-\eta_{\mu\nu}\partial_\rho\partial^\sigma {h^\rho}_\sigma + \eta_{\mu\nu} \Box
h\right)\,.
\label{eq:einstein_h}
\end{eqnarray}
Now, we have all the necessary ingredients in order to write the field Eqs. (\ref{eq:2}) in terms of the perturbation as follows 
\begin{eqnarray} 
F^{\prime} \left( R_{\mu\nu}^{(1)} - \frac{1}{2} g_{\mu\nu} R^{(1)} \right) 
+ \frac{1}{2}g_{\mu\nu} \left( F(T) - F^{\prime} T \right) 
  = {\kappa}^2 
{\mathcal{T}^{(\mathrm{M})}}_{\mu\nu}\,, 
\label{eq:2pert}
\end{eqnarray}
where we have discarded the terms higher than first order in metric quantities and considered $F(T)$ as follows. In fact, 
we suppose that the  $F(T)$  Lagrangian 
in Eq.~(\ref{eq:1}) is analytic 
 ({\it i.e.},  Taylor expandable) in term of $T$, which means that 
\begin{eqnarray}\label{sertay}
F(T)=\sum_{n}\frac{F^n(T_0)}{n!}(T-T_0)^n\simeq
F_0+F'_0T+\frac{1}{2}F''_0T^2+...\,,
\end{eqnarray}
so that the gravitational field Eqs.~(\ref{eq:2pert}) can be expressed as 
\begin{eqnarray} 
\left(F_0^{\prime}+T F_0^{\prime\prime} \right) \left( R_{\mu\nu}^{(1)} - \frac{1}{2} g_{\mu\nu} R^{(1)} \right) 
+ \frac{1}{2}g_{\mu\nu} \left[ \left(T\,F_0^{\prime}+\frac{T}{2}F_0^{\prime\prime}\right) -\left( F_0^{\prime} T+TF_0^{\prime\prime}\right)T \right]
  = {\kappa}^2 
{\mathcal{T}}^{(0)(\mathrm{M})}_{\mu\nu}\,. \nonumber\\
\label{eq:2pert1}
\end{eqnarray}
Here, ${\mathcal{T}}^{(0)(\mathrm{M})}_{\mu\nu}$ is fixed at the zeroth-order 
in Eq.~(\ref{eq:2pert1}) because, in this perturbation scheme, the first order 
on the Minkowski space has to be connected with the zeroth order of the 
standard energy-momentum tensor of matter. 
Finally, after some simplification, we obtain
\begin{eqnarray} 
F_0^{\prime}\left[ R_{\mu\nu}^{(1)} - \frac{1}{2} g_{\mu\nu} R^{(1)} \right] 
+  TF_0^{\prime\prime} \left[ R_{\mu\nu}^{(1)} - \frac{1}{2} g_{\mu\nu} R^{(1)} \right] -\frac{1}{4}T^2F_0^{\prime\prime} 
  = {\kappa}^2 
{\mathcal{T}}^{(0)(M)}_{\mu\nu}\,. 
\label{eq:2pert1s}
\end{eqnarray}
At this point, 
we can make a further assumption on $T$ that, in our case, 
due to the above-mentioned  relation $R = -T - 2{\nabla}^{\mu} \left( T^{\nu}_{\verb| |\mu\nu} \right)$, becomes $R=-T$ because the second terms is of higher order. 
The reason why we have truncated the expansion of $F(T)$  
at the second order in terms of $T$ is that we here study the 
weak field region. Since the absolute value of the torsion scalar $T$ 
is basically related to that of the curvature shown above, 
when we examine the weak field region, it is considered that the higher order terms in $T$ can be neglected. In other words, even if we include the 
higher order terms in $T$, for instance, those in proportional to $T^3$, 
the qualitative consequences would not be changed. 
In this way, the gravitational field equation is assumed to be 
\begin{eqnarray} 
F_0^{\prime}\left[ R_{\mu\nu}^{(1)} - \frac{1}{2} g_{\mu\nu} R^{(1)} \right] 
+  R^{(1)}F_0^{\prime\prime} \left[ R_{\mu\nu}^{(1)} - \frac{1}{2} g_{\mu\nu} R^{(1)} \right] -\frac{1}{4}(R^{(1)})^2F_0^{\prime\prime} 
  = {\kappa}^2 
{\mathcal{T}}^{(0)(M)}_{\mu\nu}\,, 
\label{eq:2pert1s}
\end{eqnarray}
that in terms of perturbation becomes
\begin{eqnarray} 
&& \frac{1}{2}F_0^{\prime}\left(\partial_\rho\partial_\nu {h^\rho}_\mu + \partial^\rho
\partial_\mu h_{\nu\rho} - \Box h_{\mu\nu} - \partial_\mu\partial_\nu h
-\eta_{\mu\nu}\partial_\rho\partial^\sigma {h^\rho}_\sigma + \eta_{\mu\nu} \Box
h\right) \nonumber \\ 
&& 
{}+F_0^{\prime\prime}\left(\partial_\rho\partial^\mu {h^\rho}_\mu - \Box h\right)\left(\partial_\rho\partial_\nu {h^\rho}_\mu + \partial^\rho
\partial_\mu h_{\nu\rho} - \Box h_{\mu\nu} - \partial_\mu\partial_\nu h
-\eta_{\mu\nu}\partial_\rho\partial^\sigma {h^\rho}_\sigma + \eta_{\mu\nu} \Box
h\right) \nonumber \\ 
&&
{}-\frac{1}{4}F_0^{\prime\prime} \left(\partial_\rho\partial^\mu {h^\rho}_\mu - \Box h\right)^2\,=\,{\kappa}^2 
{\mathcal{T}}^{(0)(M)}_{\mu\nu}\,. 
\label{FEh}
\end{eqnarray}
This expression can be cleaned up significantly using the {\it trace-reversed} 
perturbation $\displaystyle{\bar h_{\mu\nu} = h_{\mu\nu} - \frac{1}{2}\eta_{\mu\nu} h}$,  where $\displaystyle{\bar {h}^\mu_\mu = -h}$. 
Replacing $h_{\mu\nu}$ with 
$\displaystyle{\bar h_{\mu\nu} + \frac{1}{2}\eta_{\mu\nu} h}$ in 
Eq.~(\ref{FEh}) and expanding the equation, 
we find that all the terms with the trace $h$ are canceled. 
As a result, what remains is 
\begin{eqnarray} 
&& \frac{1}{2}F_0^{\prime}\left(\partial_\sigma\partial_\nu {{\bar h}^\rho}_{\ \mu} +
\partial^\rho \partial_\mu \bar h_{\mu\nu} - \Box \bar h_{\mu\nu} -\eta_{\mu\nu}
\partial_\rho\partial^\sigma {{\bar h}^\rho}_{\ \sigma}\right)\,=\,{\kappa}^2 
{\mathcal{T}}^{(0)(M)}_{\mu\nu}\,.\label{FEhs}
\end{eqnarray}
Applying the Lorentz gauge condition $\partial^\mu \bar h_{\mu\nu} = 0$ to the above expression, we see that all but one term vanishes:
\begin{eqnarray} 
 -\frac{F_0^{\prime}}{2}\Box \bar h_{\mu\nu}\,=\,{\kappa}^2 
{\mathcal{T}}^{(M)}_{\mu\nu}\,.\label{GW}
\end{eqnarray}
Thus, in the Lorentz gauge, the gravitational field equation for $F(T)$ 
gravity is simply reduced to the wave operator acting on the trace reversed 
metric perturbation (up to a factor $-\displaystyle{\frac{F'_0}{2 }}$) 
as in General Relativity. 
Therefore, the linearized field equation reads
\begin{equation}
\Box \bar h_{\mu\nu} = - \frac{16 \pi}{F^{\prime}_0} {\mathcal{T}}^{(M)}_{\mu\nu}\,.
\label{eq:elin}
\end{equation}
In vacuum, this equation reduces to
\begin{equation}
\Box \bar h_{\mu\nu} = 0\,.
\label{eq:elin1}
\end{equation}
Like in General Relativity, Eq.~(\ref{eq:elin}) admits a 
class of homogeneous solutions which are superpositions of plane 
waves, that is 
\begin{displaymath}
{\bar h}_{\mu\nu}({\bf x},t) = {\rm Re} \int d^3 k \ A_{\mu\nu}({\bf k}) e^{i
  ({\bf k} \cdot {\bf x} - \omega t)}\,, 
\label{eq:planewaves}
\end{displaymath}
with $\omega = |{\bf k}|$. The complex coefficients $A_{\mu\nu}({\bf
k})$ depend on the wavevector ${\bf k}$, but are independent of ${\bf
x}$ and $t$. They are subject to the constraint $k^\mu A_{\mu\nu} = 0$
(which follows from the Lorentz gauge condition) with $k^\mu =
(\omega,{\bf k})$, but are otherwise arbitrary. These solutions are the 
gravitational waves.

{}From this result, it is clear that $F(T)$ cannot be a ``signature'' to discriminate further gravitational wave modes or polarizations at the first order in linearized theory. It is important to note that this result is completely different from that in $F(R)$ gravity, where it is evident that there exist further degrees of freedom of the gravitational field~\cite{SCF}. In particular, 
it can be found that besides a massless spin-2 field (the standard graviton), 
$F(R)$ gravity theories contain also spin-0 and spin-2 massive modes with the latter being, in general, ghost modes. As shown in Ref.~\cite{Li:2011rn}, there are $D-1$ extra degrees of freedom for $F(T)$ gravity in $D$ dimensions. This implies that the extra degrees of freedom correspond to one massive vector field or one massless vector field with one scalar field. These modes do not contribute to the gravitational radiation if it considered, as standard, at first-order perturbation theory.

\textit{Scalar-tensor representation of $F(T)$ gravity}--
Let us now consider, by analogy to $F(R)$ gravity, the scalar-tensor representation of $F(T)$ gravity, where it is straightforward to check that the scalar mode does not propagate unlike the case of $F(R)$ gravity. The action for $F(T)$ gravity can be written as follows
\be
S=\int d^4x\ |h| \left(\phi T-V(\phi)\right)\ .
\label{ST1}
\ee
By varying the action with respect to $\phi$, the correspondence between the action (\ref{ST1}) and that of $F(T)$ gravity is found as 
\be
V'(\phi)=T  \rightarrow  \phi=\phi(T) \,, 
\label{ST2}
\ee
namely, in principle, 
we can solve the first equation in (\ref{ST2}) in terms of $\phi$ 
as the second equation. 
This yields 
\be
F(T)= \phi(T) T- V(\phi(T))\ . 
\label{ST3}
\ee
Moreover, the scalar field and its potential can be written in terms of the function $F(T)$ as 
\be
\phi=F^{\prime}(T)\ , \quad V(\phi)=F^{\prime}(T)\ T-F(T)\ .
\label{ST3a}
\ee
On the other hand, 
the gravitational field equations can easily be obtained by the variation of 
the action (\ref{ST1}) with respect to the tetrad $h^{a}_{\;\;\mu}$ as
\be
\left[ 
\partial_\mu \left( h h_A^\nu S_\nu^{\verb| |\lambda\rho} \right) 
-h h_A^\rho S^{\mu\nu\lambda} T_{\mu\nu\rho} \right] \phi 
+h h_A^\nu S_\nu^{\verb| |\lambda\rho} \partial_\rho \phi 
+\left(1/2\right) h h_A^\lambda \left(\phi T-V(\phi)\right) = {\kappa}^2 
{\mathcal{T}^{(\mathrm{M})}}_A^\lambda\ . 
\label{ST4}
\ee
In a covariant form, the gravitational field equation (\ref{ST4}) reads
\be
\left( R_{\mu\nu} - \frac{1}{2} g_{\mu\nu} R \right) \phi 
- \frac{1}{2}g_{\mu\nu} V(\phi)
+  S_{\nu\mu\rho} \nabla^{\rho} \phi  = {\kappa}^2 
{\mathcal{T}^{(\mathrm{M})}}_{\mu\nu}\,. 
\label{ST5}
\ee
Thus, as executed above, we may explore the weak field limit by providing that 
$g_{\mu\nu}\,=\,\eta_{\mu\nu}+h_{\mu\nu}$, and that the scalar field can be described by a constant background value $\phi_0$ 
plus a small perturbation $\delta\phi$ around it as 
\be
\phi=\phi_0+\delta\phi\ . 
\label{ST6}
\ee 
Furthermore, 
the scalar potential can be expanded in powers of the perturbations as
\be
V(\phi)=V_0+V_0^{\prime}\delta\phi+{\cal O}(\delta\phi^2)\ .
\label{ST7}
\ee
Accordingly, the gravitational field equation (\ref{ST5}) at the first order of the perturbations becomes 
\be
\left( R_{\mu\nu}^{(1)} - \frac{1}{2} \eta_{\mu\nu} R^{(1)} \right) \phi_0 
- \frac{1}{2}\left(h_{\mu\nu} V_0+\eta_{\mu\nu}V_0^{\prime}\delta\phi\right)  = {\kappa}^2 
{\mathcal{T}^{(\mathrm{M})}}_{\mu\nu}\,. 
\label{ST8}
\ee
Note that the tensor 
$S_\rho^{\verb| |\mu\nu}$ consists of 
only first derivatives of the tetrads, and therefore it becomes null at the zeroth order and the last term in the left-hand side (l.h.s.) of Eq.~(\ref{ST5}) is null at the first order in the perturbations. Moreover, the scalar torsion $T$ also becomes null at the zeroth order, and by the scalar field equation (\ref{ST2}), the derivative of the scalar potential evaluated at $\phi = \phi_0$ is given by 
\be
V_0^{\prime}=T_0=0\ ,
\label{ST9}
\ee
where $T_0$ is the value of $T$ at $\phi = \phi_0$, 
so that Eq.~(\ref{ST8}) can turn out 
\be
R_{\mu\nu}^{(1)} - \frac{1}{2} \eta_{\mu\nu} R^{(1)}
+h_{\mu\nu} \Lambda  = \frac{{\kappa}^2}{\phi_0} 
{\mathcal{T}^{(\mathrm{M})}}_{\mu\nu}\ . 
\label{ST10}
\ee
This coincides with the Einstein equation at the first order of the perturbations in the presence of a cosmological constant $\Lambda=-\frac{1}{2}V_0=\frac{1}{2}F(T = 0)$, and hence the well known result of General Relativity for gravitational waves is recovered. Consequently, in $F(T)$ gravity, unlike in $F(R)$ gravity, 
there is no propagating scalar modes in the gravitational waves at least 
when a flat background is assumed.

\textit{Summary}-- 
We have investigated gravitational waves in $F(T)$ gravity. 
In particular, in the Minkowskian limit for a class of analytic function $F(T)$ in the Lagrangian, 
we have explicitly shown that gravitational wave modes in $F(T)$ gravity are 
the same as those in General Relativity. 
By using this representation, 
it has been shown that the scalar field does not propagate at the first order of the perturbations, because the only remaining terms of the scalar field in the perturbed equations are the zeroth order, so that the Einstein equation in General Relativity with a cosmological constant proportional to $F(T = 0)$ can be recovered. It should be emphasized that the cosmological constant can exist 
only if $F(T = 0) \neq 0$.


\textit{Acknowledgments}-- 
S.C.,  D.S-G., and M.D.L.  sincerely acknowledge the KMI visitor program at 
Kobayashi-Maskawa Institute for the Origin of Particles and the Universe (KMI), Nagoya University, 
where this work has been initiated. 
Furthermore, we express our sincere gratitude 
to Professor Sergei D. Odintsov for continuous encouragements and 
various useful discussions. 
The work is supported in part by the JSPS Grant-in-Aid for 
Young Scientists (B) \# 25800136 (K.B.) 
and 
that for Scientific Research 
(S) \# 22224003 and (C) \# 23540296 (S.N.). 
In addition, D.S-G appreciates the support from the University of the Basque Country and the URC financial support from the University of Cape Town (South Africa).   S.C. and M.D.L. acknowledge the support of INFN Sez. di Napoli (Iniziative
Specifiche NA12 and OG51).


\end{document}